\documentclass[12pt]{article}
\usepackage[english]{babel}
\usepackage[latin1]{inputenc}
\usepackage{pgf}
\usepackage{graphicx}
\usepackage{times}
\usepackage[T1]{fontenc}
\usepackage{epsfig}
\usepackage{amsmath}
\usepackage{amssymb}
\begin{document}

\title{Three-body forces from a classical nonlinear
field }

\author{H. Arod\'z$^a$, J. Karkowski$^a$ and Z.
\'Swierczy\'nski$^b$ \\$\;\;$ \\ \emph{\small $^a$
Institute of Physics,
Jagiellonian University, Cracow, Poland }\\ \emph{\small
$^b$Institute of Computer Science and Computer Methods,}\\\emph{\small Pedagogical University, Cracow, Poland}}

\date{$\;$}

\maketitle

\begin{abstract}
Forces in the systems of two opposite sign and three identical
charges coupled to the dynamical scalar field of the signum-Gordon
model are investigated. Three-body force is present, and the
exact formula for it is found. Flipping the sign of one of the
two charges changes not only the sign but also the magnitude of
the force. Both effects are due to nonlinearity of the field
equation.
\end{abstract}

\vspace*{2cm} \noindent PACS: 11.27.+d, 11.10.Lm, 03.50.Kk
\\

\pagebreak

\section{ Introduction}

As is well-known, relativistic scalar fields play the crucial roles
in physics of fundamental interactions from particle physics to
cosmology. Topological and non-topological solitons \cite{1},
\cite{2}, long lived oscillons \cite{3}, phenomena such as
spontaneous symmetry breaking and Higgs mechanism \cite{4}, all
can hardly be considered without scalar fields. Not surprisingly,
one can find in literature a whole variety of field-theoretic
models with scalar fields. We have been interested in the so
called signum-Gordon model which involves just one classical
scalar field $\varphi$, real or complex. Its defining feature is
the V-shaped self-interaction potential proportional to the
modulus of the field, $U(\varphi) = g |\varphi|$, where $g>0$ is
the self-coupling constant. This model originated from
investigations of perturbations of the ground state of a system of harmonically  coupled
pendulums bouncing from a stiff rod in the constant gravitational
field \cite{5}. It has turned out that it has truly amazing
properties. Its basic dynamical features, which do not depend on
the dimension of space-time, include a scale invariance of the
on-shell type, and generically a very fast, parabolic approach of
the field to its ground state value $\varphi=0$, which is reached
exactly on a finite distance. Because of the latter property, the
signum-Gordon field may formally be regarded as an ultramassive one,
because the massless or massive fields have a different asymptotic
behavior of  Coulomb or Yukawa type, respectively.

 Furthermore, in the case of real signum-Gordon field the
field equation has the form
\begin{equation}
\partial_{\mu} \partial^{\mu} \varphi = - g \:
\mbox{sign}\:\varphi,
\end{equation}
where the $\mbox{sign}$ function has the standard values $\pm 1$
and $\mbox{sign}\:0=0$. Thus, the r.h.s. of this equation is
piecewise constant, greatly facilitating construction of
interesting analytic solutions. In fact, its solutions include
non-radiating oscillons \cite{6}, as well as a whole family of
self-similar fields \cite{7}. In the case of complex scalar field
with the V-shaped self-interaction compact Q-balls were found
\cite{8}. In all these cases pertinent exact analytic solutions
were  obtained. Thus  the  signum-Gordon model  has turned out to
be a very good theoretical laboratory for studying the highly
nontrivial, nonlinear dynamics of scalar fields.

Recently, we have investigated forces exerted on static external
charges interacting with the real signum-Gordon field \cite{9}. In
particular, the force $F_{q q}(a)$  between two identical,
separated by the distance $a$, point static charges of the
strength $q$  has been calculated.  It exactly vanishes when the
distance between the charges exceeds certain finite value $a_{*}$
that depends on their strengths.  Such unusual behavior is due to
the fact that $U'= g \:\mbox{sign}\:\varphi$  remains finite even
for arbitrarily  small values of the field $\varphi$. This is
similar to the constant gravity force  attracting a ball to floor.
The scalar field forms a compact cloud surrounding the charges. The
force is attractive.

In the present paper we extend the work  \cite{9} by  presenting
two effects that are  due to nonlinearity of the signum-Gordon
field. The point is that in the case of one-dimensional space one
can construct the exact solutions of the inhomogeneous
signum-Gordon equation (Eq.\ (4) below) for any number of static,
point-like external sources. This gives us  the rare opportunity of 
having  full  knowledge  about  the effects that are due to the
self-interaction of the signum-Gordon field -- the mediating field
for the forces between the external charges. Specifically, we
consider the force $F_{-q q}(a)$ between two opposite static
charges $-q$ and $q$ separated by the distance $a$,  and  the
forces in the system of three identical static charges. We 
obtain exact formulas for the scalar mediating field and for the
forces. In the $-q q$ case, the field and force vanish when the
distance $a$ exceeds the same finite value $a_{*}$ as in the $q q$
system investigated in \cite{9}.  The  force is repulsive, as
expected. Our
 new finding is that the magnitudes of the forces in the $qq$ and $-qq$
 cases  are different, $|F_{-q q}(a)| \neq |
F_{q}(a)|$, as opposed to the case of charges coupled to a free
field, e.g., electric charges  interacting with the electromagnetic field. Our
most interesting result however is the observation that in the case of three
charges a three-body force is present. It is clear that in general
one should expect $N$-body forces when there are $N$ charges. These
effects are due to the nonlinearity of field equation and
therefore are expected to appear in other nonlinear
field-theoretic models.

The plan of our paper is as follows. In the next section we
briefly recall the method of calculating the forces, we explain
how one can obtain the pertinent solution of field equation, and
we discuss the $-q q$ case. The three-body forces in the $q q q$
case are calculated in Section 3. Summary and remarks are
collected in Section 4.

\section{The force in the $-qq$ case }

We consider  two static, point-like charges of the opposite sign
interacting with the dynamical real scalar field $\varphi$ .  The
Lagrangian for this system has the form (we use the $c=\hbar =1$
units)
\begin{equation} L = \frac{1}{2} \partial_{\mu} \varphi
\partial^{\mu} \varphi
 - g \:|\varphi|   + j \varphi, \end{equation} where
\begin{equation}
j(x) = q \: \delta (x-a) - q \: \delta (x).
\end{equation}
Here $\mu =0, 1$, and $x\equiv x^1$ is the spatial coordinate in the one-dimensional space.
The charges are located at  $x=0$ and $x=a>0$.
The field equation corresponding to this Lagrangian reads
\begin{equation}
\partial_{\mu} \partial^{\mu} \varphi + g\: \mbox{sign}\: \varphi  = j(x).
\end{equation}
The simplest way to obtain the $\mbox{sign} \:\varphi $ term is
first to regularize the  field potential, e.g.,  $\:g\:|\varphi|
\rightarrow g\: \sqrt{\delta^2 + \varphi^2}$, and to take the
limit $\delta \rightarrow 0$ in the term  $g\: \varphi/
\sqrt{\delta^2 + \varphi^2}$  that appears in  the corresponding
Euler-Lagrange equation.
 Thus,  \[ \mbox{sign}\: \varphi  = \lim_{\delta \rightarrow 0} \frac{\varphi}{\sqrt{\delta^2 + \varphi^2}}.  \]
It is clear from this definition that  $ \mbox{sign} \:0 =0$. Such
regularized version of the signum-Gordon model, i.e.,  with
$\delta
>0$, was investigated on its own right  in \cite{10}.

The force exerted on the charge $-q$  located at $x=0$  is given
by the total flux of  momentum towards that charge,  as explained
in, e.g.,    \cite{11}, \cite{9}.    The  momentum density and the
flux of momentum are given by the energy-momentum tensor of our
system,
\begin{equation}
T_{\mu\nu} = \partial_{\mu} \varphi \partial_{\nu} \varphi
-
\eta_{\mu\nu} L,
\end{equation}
where $(\eta_{\mu\nu}) = \mbox{diag}(1, -1)$ is the
Minkowski  metric.

The total momentum $P^1$ of the field,  given by
\[
P^1 = - \int dx \:T_{01} = - \int dx \:\partial_0 \varphi \:
\partial_x \varphi,
\]
vanishes in the case of static fields we consider.  The presence
of the point charges  breaks the translational symmetry. In
consequence, instead of the continuity equation  $
\partial^{\mu} T_{\mu\nu} =0$  we  have
\begin{equation}
 \partial^{\mu} T_{\mu\nu} = -  \varphi \partial_{\nu} j.
\end{equation}

The flux of momentum along the $x$ axis  is given by $T_{11} =
(\partial_x \varphi)^2/2 - g \: |\varphi|$ if $x \neq 0, a$.  By the
definition, the force exerted on the charge located at $x=0$ is
given by the rate at which the momentum is transferred to that
charge.  Hence,
\begin{equation} F^1 = \left. T_{11}\right|_{x= -\epsilon} -
\left.T_{11}\right|_{x=\epsilon}. \end{equation} Here $\epsilon$
can be any number from the interval $(0, a)$, because  in the
static case the identity (6) implies that $T_{11}$ does not depend
on $x$ in the open intervals $(-\infty,0)$, $(0,a)$, $(a,
\infty)$, in which $j=0$. Furthermore, we will see that the
pertinent solution $\varphi$ and $T_{11}$ vanish in the interval
$(-\infty, -d_0)$, where $d_0>0$ is a constant. Therefore, $
\left. T_{11}\right|_{x= -\epsilon} =0$, and formula (7) is
simplified to
\begin{equation}
F^1 = - \left.T_{11}\right|_{x=\epsilon}.
\end{equation}

It remains to find the field $\varphi$  in the presence of the
sources. In the static case it satisfies the equation
\begin{equation}
\partial^2_x \varphi - g \: \mbox{sign}\: \varphi = q \delta(x) - q
\delta(x-a).
\end{equation}
In order to ensure finiteness of the total energy $E= \int dx \:
T_{00}$, where $T_{00} = (\partial_x \varphi)^2/2 + g \:
|\varphi|$, the field $\varphi$ and $\partial_x \varphi$  should
vanish for $|x| \rightarrow \infty$. Furthermore, integrating both
sides of Eq.\ (9) over intervals of the half-length $\delta <a$
around $x=0$ and $x=a$ we obtain the conditions
\begin{equation}
\partial_x \varphi(\delta) - \partial_x \varphi(-\delta)=q,
\;\;\;\;
\partial_x \varphi(a+\delta) - \partial_x \varphi(a-\delta)=-q,
\end{equation}
which are the one-dimensional counterpart of Gauss law of
electrostatics. These conditions imply that $\partial_x \varphi$
is discontinuous at the  locations of the charges.

Obvious approach  to solving Eq.\ (9) is first to find solutions
of the homogeneous equation
\begin{equation}
\partial^2_x \varphi - g \: \mbox{sign}\: \varphi = 0
\end{equation}
in the open intervals  $(-\infty,0),$ $(0,a)$ and  $(a, \infty)$.
Next, we glue them at $x=0$, $x=a$, so that $\varphi$ is
continuous function of $x$ at these points, and that the
conditions (10), in which we may take the limit $\delta \rightarrow 0+$, are
satisfied.

The general solution of Eq.\ (11) in the interval in which
$\varphi >0$, i.e., $\mbox{sign}\: \varphi = +1$, has the form
\begin{equation}
\varphi(x) = \frac{g}{2} x^2 + Ax +B,
\end{equation}
and if  $\mbox{sign} \: \varphi = -1$ then
\begin{equation}
\varphi(x) = - \frac{g}{2} x^2 + Cx +D,
\end{equation}
where $A, B, C, D$ are constants to be determined from the
matching conditions.  There also exists the trivial solution
$\varphi=0$ which represents the ground state of the classical
signum-Gordon field. Our Ansatz for the solution has the following
form
\begin{equation} \varphi_{-qq}(x) = \left\{ \begin{array}{ll}
0 & \;\;\; x \leq -d_0, \\ u_1(x)& \;\;\; x\in [- d_0, 0],
\\ u_2(x)  & \;\;\; x \in [0, d_1], \\ u_3(x)  & \;\;\; x \in [d_1, a], \\
u_4(x)  & \;\;\; x \in [a, a+d_2],\\ 0 & \;\;\; x \geq a+d_2,
\end{array} \right.
\end{equation}
where $d_0>0$, $d_2>0$, $d_1\in(0,a)$  are constants to be
determined. The functions $u_1, u_2$ are negative inside their
domains and have the form (13), while $u_3, u_4$ are positive and
have the form (12), see Fig.\ 1.

\begin{center}
\begin{figure}[tph!]
\hspace*{1cm}
\includegraphics[height=3cm, width=12cm]{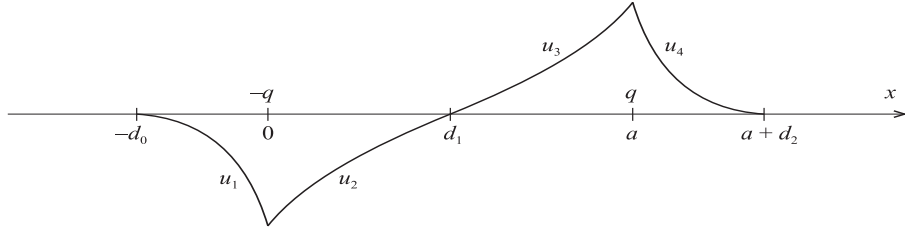}
\caption{Schematic picture of the nontrivial part of the solution
(14) for the $-qq$ case }
\end{figure}
\end{center}

The function $u_1$ matches the ground state solution $\varphi=0$
at the point $x=-d_0$. The matching conditions are the continuity
of $\varphi$ and $\partial_x\varphi$. Trivial calculation gives
\begin{equation}
u_1(x) = - \frac{g}{2} \left(x+d_0\right)^2.
\end{equation}

The function $u_2$ obeys the conditions
\[ u_2(0) = u_1(0), \;\;\; \partial_x u_2(0) - \partial_xu_1(0)=q,
\;\;\;u_2(d_1)=0.\] Simple calculations give
\begin{equation}
u_2(x) = - \frac{g}{2} (d_1-x) \left(\frac{d_0^2}{d_1} -x\right),
\end{equation}
and the relation
\begin{equation} (d_0 +d_1)^2 = \frac{2 q}{g}
d_1.
\end{equation}

The function $u_4$ has the general form (12) with the coefficients
$A,B$ determined from the conditions  $u_4(a+d_2)=0$, $\:
\partial_x u_4(a+d_2)=0$, which ensure the correct matching of
$u_4$ with the trivial solution $\varphi=0$  at the point
$x=a+d_2$. It turns out that
\begin{equation}
u_4(x) =  \frac{g}{2} \left(x-a - d_2\right)^2.
\end{equation}

The function $u_3$ of the general form (12) has to obey four
conditions: the matching conditions with the function $u_2$ at the
point $x=d_1$, i.e., $u_3(d_1)=0$, $ \partial_x u_2(d_1) =
\partial_x u_3(d_1)$, and  the conditions
$u_3(a)=u_4(a)$, $\partial_x u_3(a) - \partial_x u_4(a)=q$ at the point $x=a$. These
conditions determine the precise form of $u_3$, namely
\begin{equation}
u_3(x)= \frac{g}{2} (x-d_1) \left(x-2d_1 +
\frac{d_0^2}{d_1}\right),
\end{equation}
and also give the following relations
\begin{equation}
(d_2+a-d_1)^2 = \frac{2q}{g}(a-d_1), \;\;\;\; 2 d_1+d_0 = a+d_2.
\end{equation}
These relations together with (17) fix the constants $d_0, d_1,
d_2$:
\begin{equation}
d_0=d_2 = \sqrt{\frac{qa}{g}} - \frac{a}{2}, \;\;\;\; d_1 =
\frac{a}{2}.
\end{equation}

The force exerted on the charge $-q$ located at $x=0$ is given by
formula (8), in which we put $T_{11}|_{x=\epsilon} = (\partial_x
u_2(0))^2/2 - g u_2(0)$.  Simple calculation gives
\begin{equation}
F_{-qq}^1(a) = - \frac{1}{2} q^2\left( 1 -
\sqrt{\frac{a}{a_*}}\right)^2,
\end{equation}
where \[ a_* = \frac{q}{g}.\] We see that the force is repulsive
one. It contains the one-dimensional scalar Coulomb force $F^1_{Coul}=
-q^2/2$ as the leading term when $a \ll a_*$.

The solution (14) and formula (22) for the force are valid when $a
\in (0, a_*]$. When the distance $a$ between the charges is equal
to $a_*$ the charges become completely  screened by the scalar
field. When $a > a_*$, each charge  is surrounded by a compact
cloud of the field of the width $a_*$. The clouds do not overlap
-- in between them there is the region with the ground state field
$\varphi=0$. Thus the charges do not feel the presence of each other,
and  the force vanishes of course. Such screening has been
observed already in \cite{9} in the case of two identical
point-like charges $qq$ located at $x=0$ and $x=a$. In this case
the force exerted on the charge located at $x=0$ is given by the
following formula
\begin{equation}
F^1_{qq}(a) = \frac{1}{2} q^2 \left(1- \frac{a}{a_*}\right).
\end{equation}
Apart from the difference in sign, which means that $F^1_{qq}(a)$
is an attractive force, we see different dependence on the
distance $a$. Because this difference disappears when we put the
coupling constant $g=0$, it is related to the presence of  the
nonlinear $\mbox{sign} \:\varphi$ term in the field equation (4).

\section{The three-body forces}

Let us now consider three identical point charges of the strength
$q$ located at the points $x=-a$, $x=0$ and $x=b$, where $a, b
>0$.  In this case  Eq.\ (4) takes the form \begin{equation}
\partial^2_x \varphi - g \: \mbox{sign} \: \varphi = q \delta(x+a) + q \delta(x) + q
\delta(x-b).
\end{equation}
The Ansatz for the solution has the form
\begin{equation} \varphi_{qqq}(x) = \left\{ \begin{array}{ll}
0 & \;\;\; x \leq -a-d_-, \\ w_1(x)& \;\;\; x\in [-a - d_-, -a],
\\ w_2(x)  & \;\;\; x \in [-a, 0], \\ w_3(x)  & \;\;\; x \in [0, b], \\
w_4(x)  & \;\;\; x \in [b, b+d_+],\\ 0 & \;\;\; x \geq b+d_+,
\end{array} \right.
\end{equation}
where $d_+, d_->0$ are constants. All functions $w_i(x)$,
$i=1,..,4$, are strictly positive inside their domains,  so they
have the general form (12). The nontrivial part of the function $\varphi_{qqq}(x)$ is
depicted  in Fig.\ 2.

\begin{center}
\begin{figure}[tph!]
\hspace*{1cm}
\includegraphics[height=3.0cm, width=12cm]{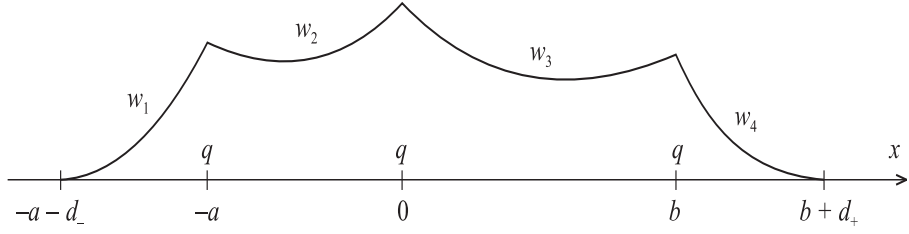}
\caption{Schematic picture of the nontrivial part of the solution
(25) for the case of three identical charges}
\end{figure}
\end{center}

The constants in these functions and the constants $d_{\pm}$ are
determined  from the matching conditions which ensure continuity
of $\varphi_{qqq}$  everywhere and continuity of $\partial_x \varphi_{qqq}$
at $x=-a-d_- $, $x= b+d_+$, and from the `Gauss law' relations of
the type (10) at $x=-a$, $x=0$, $x=b$. Because the pertinent
calculations are straightforward and completely analogous to the
ones presented in the -qq case, we just quote the  results:
\begin{equation}
w_1(x)=\frac{1}{2}\: g \left(x+a+d_-\right)^2,
\end{equation}
\begin{equation}
w_2(x)=\frac{1}{2}\: g \left(x+a+d_-\right)^2 - q (x+a),
\end{equation}
\begin{equation}
w_3(x)=\frac{1}{2}\: g \left(x-b-d_+\right)^2 + q (x-b),
\end{equation}
\begin{equation}
w_4(x)=\frac{1}{2}\: g \left(x-b-d_+\right)^2,
\end{equation}
where
\begin{equation}
d_-= \frac{3}{2}\: a_* - \frac{1}{3}\: b - \frac{2}{3}\: a, \;\;\;
d_+= \frac{3}{2}\: a_* - \frac{2}{3}\: b - \frac{1}{3}\: a.
\end{equation}

The solution (25) is correct provided that $a$ and $b$ are not too
large, otherwise one charge (or more) will be completely screened
 by the compact cloud of the scalar field, as discussed in \cite{9}.  Such
 a screened charge  decouples from the others.  In such cases,
 in between the clouds there are segments of the $x$ axis where
 the field has its ground state value $\varphi=0$, and the force
 on such a distant charge exactly vanishes.  More precisely,   for the validity  of the solution (25) $a$
 and $b$ should  satisfy the following bounds
 \begin{equation}
 2 a + b < 3 a_*, \;\;\;\; a + 2 b < 3 a_*.
 \end{equation}
The decoupling of the charge located at $x=-a$ from the two other
occurs when $a=3a_*/2 - b/2$. This can be found by checking when
the function $w_2$ has a zero in the interval $(-a, 0)$. If $w_2
<0$ in that interval, the function (25) ceases to be the solution
of Eq.\ (24). The decoupling of the charge located at $x=b$ takes
place when $b=3 a_*/2 - a/2$. In this case one should check
positivity of the $w_3$ function.

It is clear from the  considerations at the beginning of Section 2 that the total force exerted on the charge located at $x=-a$ is given by the formula
\begin{equation}
F_{-a} = - \left.T_{11}(w_2)\right|_{x=-a} = q^2 \left(1 - \frac{2a+b}{3 a_*}\right).
\end{equation}
This force is the attractive one.

Let us compare this force with the sum of two-body forces exerted on this
charge by the charges located at $x=0$ and $x=b$.  For the two-body forces we  use
formula (23) assuming that $a<a_*$, $a+b < a_*$, so that our charge is in the
interaction range of the two other charges.  We obtain
\[ F_{2body}(-a) = q^2 \left(1 - \frac{2a+b}{2 a_*}\right). \]
The difference
\[  F_{3body}(-a) = F_{-a} - F_{2body}(-a)  = q^2\frac{2a+b}{6 a_*}  \]
gives the three-body component of the total force  exerted on that charge.
This component is negligibly small if  $a \ll a_*$, $b \ll a_*$ (when the constant scalar Coulomb
force dominates), but it cannot be neglected if the distances between the charges become  comparable with $a_*$.

The  force  exerted on the charge  located at $x=0$ also has a three-body component.
The total force is calculated from the formula
\[ F_0 = \left. T_{11}(w_2)\right|_{x=0} -  \left. T_{11}(w_3)\right|_{x=0} = q^2 \frac{a-b}{3 a_*}. \]
On the other hand, the sum of two-body forces  exerted by the neighboring charges is equal to
\[
 F_{2body}(0) = q^2 \frac{a-b}{2 a_*}. \]
Here we assume that $a,b \ll a_*$, so that the use of formula (23) is formally justified.  Again,
the three-body component ,
\[  F_{3body}(0) = F_{0} - F_{2body}(0)  = q^2 \frac{b - a }{6 a_*},  \]
can have a sizable magnitude.

The force exerted on the charge located at $x=b$ is given by the
formula
\[ F_b = \left.T_{11}(w_3)\right|_{x=b}= -q^2 \left(1 -
\frac{a+2b}{3 a_*}\right).  \] It differs from the force $F_{-a}$
by the sign, and $a$ and $b$ are interchanged, as  expected in
view of the spatial structure of  our $qqq$ system.

\section{Summary and remarks}

\noindent 1. We have found the exact form of the scalar field in
the presence of two and three point-like external charges in one
dimensional space. For simplicity, we have considered the $-qq$
and $qqq$  systems, but a generalization to charges of arbitrary
strength is straightforward.  The fields have the parabolic,
compact tails that are  characteristic for models with the V-shaped
self-interactions. Next, we have calculated the forces exerted on
the charges. Our main finding is that the forces are shaped mainly
by the self-coupling of the mediating  field. Only at the very short
distances ($a \ll a_{*}$) the familiar scalar Coulomb force dominates,
and the self-interaction of the  field is not important.

\noindent 2.  Particularly interesting is the presence of the
three-body force. Generalizing our result,  we expect that in
a system of $N$ particles coupled to a non-linear field all
$n$-body forces will appear with $n=2,3,\ldots N$. We have seen
that the three-body force can have a significant strength. This
would cast a shadow on attempts to model dynamics of many particle
systems by Hamiltonians that include only two-particle
interactions.  The importance of the many-body forces has recently
been emphasized in, e.g., \cite{12} in the context of nuclear
physics, and in \cite{13} in condensed matter physics.

\noindent 3.  It is clear that  nonlinear field-theoretic effects
in interactions of systems of many particles are in general
present and important.  As a good illustration of this point one
may take the results presented in   \cite{14},  where an
explanation  of rotation curves of galaxies  without invoking the
concept of dark matter  is proposed.  It would be very interesting
to investigate the many-body forces in the case of particles
coupled to a non-Abelian gauge field of the $SU(n)$ type.

\end{document}